\documentclass[a4paper,fleqn]{cas-dc}

\usepackage[authoryear,longnamesfirst]{natbib}
\usepackage[linesnumbered,ruled,vlined]{algorithm2e}
\usepackage{latexsym}
\usepackage{graphicx}
\usepackage{amssymb}
\usepackage{fontawesome}
\usepackage{tabularx}
\usepackage{utfsym}
\usepackage{CJKutf8}
\usepackage{amsmath}
\usepackage{afterpage}

\usepackage{url}

\newcommand{\secref}[1]{§\ref{#1}}

\begin{document}

\begin{CJK}{UTF8}{gbsn}
\let\WriteBookmarks\relax
\def\floatpagepagefraction{1}
\def\textpagefraction{.001}
\def \sn {OMNISEC}



\title [mode = title]{\sn{}: LLM-Driven Provenance-based Intrusion Detection via Retrieval-Augmented Behavior Prompting}

%

\author[1]{Wenrui~Cheng}



\author[1]{Tiantian~Zhu}

\fnmark[*]

\ead{ttzhu@zjut.edu.cn}
\ead[url]{<URL>}

\author[1]{Shunan~Jing}
\author[1]{Jian-Ping~Mei}
\author[1]{Mingjun~Ma}
\author[1]{Jiaobo~Jin}
\author[2]{Zhengqiu~Weng}


\affiliation[1]{organization={Zhejiang University of Technology},
country={China}}
\affiliation[2]{organization={Wenzhou University of Technology},
country={China}}


\credit{<Credit authorship details>}
\maketitle

\begin{abstract}
Recently, Provenance-based Intrusion Detection Systems (PIDSes) have been widely used for endpoint threat analysis. These studies can be broadly categorized into rule-based detection systems and learning-based detection systems. Among these, due to the evolution of attack techniques, rules cannot dynamically model all the characteristics of attackers. As a result, such systems often face false negatives. Learning-based detection systems are further divided into supervised learning and anomaly detection. The scarcity of attack samples hinders the usability and effectiveness of supervised learning-based detection systems in practical applications. Anomaly-based detection systems face a massive false positive problem because they cannot distinguish between changes in normal behavior and real attack behavior. The alert results of detection systems are closely related to the manual labor costs of subsequent security analysts. To reduce manual analysis time, we propose \sn{}, which applies large language models (LLMs) to anomaly-based intrusion detection systems via retrieval-augmented behavior prompting.

First, \sn{} uses lossless reduction technology to remove redundant information flows from the provenance graph. Second, \sn{} builds two external knowledge databases: a benign behavior knowledge base based on normal behavior activities within the host and a threat intelligence knowledge base based on external threat intelligence. \sn{} can identify abnormal nodes and corresponding abnormal events by constructing suspicious nodes and rare paths. By combining two external knowledge bases, \sn{} uses Retrieval Augmented Generation (RAG) to enable the LLM to determine whether abnormal behavior is a real attack. Finally, \sn{} can reconstruct the attack graph and restore the complete attack behavior chain of the attacker's intrusion. Experimental results show that \sn{} outperforms state-of-the-art methods on public benchmark datasets.
\end{abstract}


\section{Introduction}\label{}
In the highly interconnected digital age, Advanced Persistent Threats (APTs) have become the primary means of disrupting the security of critical infrastructure and sensitive data \cite{xing2020review}. These cyber-attack methods are characterized by high concealment, long-term persistence, clear targeting, and high modularity \cite{li2016study}. Attackers often carefully plan their attacks. They combine various methods and techniques to exploit vulnerabilities. Once inside, attackers persist undetected, stealing data, manipulating systems, or sabotaging operations, leading to severe financial and security consequences. For example, in the 2020 SolarWinds supply chain incident \cite{whitaker2021solarwinds}, attackers implanted malicious code in software updates, leading to the compromise of several US government departments and large enterprises. The attack lasted for months without being detected, causing serious national-level information leakage.


Facing the increasingly severe APT threats, Provenance-based Intrusion Detection Systems (PIDSes) have attracted widespread attention due to their advantages in detecting and tracing complex attacks \cite{zengy2022shadewatcher,hassan2019nodoze,hassan2020we,ma2016protracer}. Compared with traditional Intrusion Detection Systems (IDSes), PIDSes construct provenance graphs to represent the causal relationships between events, recording low-level behaviors such as system calls and process communications, thereby achieving precise modeling of the attack chain and improving detection accuracy and event traceability.

However, existing PIDSes still face several challenges in practical applications. Although rule-based heuristic methods \cite{milajerdi2019holmes,milajerdi2019poirot,hossain2017sleuth,hassan2020tactical} demonstrate high efficiency in detecting known malicious behaviors, they rely heavily on predefined rules and attack patterns, such as specific sequences of system calls, making it difficult for them to cope with the constantly emerging new types of attacks. Meanwhile, attackers may modify their attack strategies to evade static rules, leading to a high rate of false negatives in the system. On the other hand, supervised learning-based methods \cite{alsaheel2021atlas,zhang2019deep} categorize system behaviors into benign and malicious classes and rely on a large amount of high-quality labeled data to train models. When there is sufficient data, they can achieve good detection results. However, the coverage of the training samples often limits their generalization ability \cite{wang2021network}, and they perform poorly when facing unknown attacks. In addition, supervised learning methods require a large amount of labeled data, which is costly to obtain and difficult to adapt to the complex and changing real-world operating scenarios.

Anomaly detection methods identify potential abnormal behaviors by learning the normal behavior patterns of a system. These methods typically employ unsupervised learning to detect unknown attacks without requiring large amounts of labeled data. Currently, a significant amount of work has been done to build intrusion detection systems based on anomaly detection \cite{wang2022threatrace,yang2023prographer,cheng2024kairos,han2020unicorn,manzoor2016fast,xie2018pagoda,wang2020you}. However, anomaly detection essentially uses “deviation from normal” as the sole criterion and is trained solely on benign behavior data \cite{zhang2009survey}. This makes it difficult for the model to accurately distinguish between normal variations and actual attack behaviors, leading to a high rate of false positives when encountering new datasets. Moreover, this results in a large number of alerts in practical deployment, which still require manual filtering and confirmation, significantly increasing the cost of security. 
In an enterprise environment, anomaly detection models are usually trained on benign operation logs from hosts to learn their stable behavior patterns. 
When the user of the host changes, the behavior pattern may change as well. For example, a new employee may take over and perform operations that differ significantly from previous ones. These operations might still be legitimate. However, the model has only learned the existing behavior patterns. As a result, it may misjudge these new operations as anomalies, which further increases the false positive rate.

Detection methods that rely solely on a single knowledge source (such as normal behavior patterns) can identify abnormal behavior to some extent. However, since they do not combine other types of knowledge (such as known attack patterns or threat intelligence), they often fail to accurately distinguish between malicious and non-malicious anomalies when facing complex scenarios or changes in strategies \cite{ben2015effects} (\textbf{Challenge 1}). Some anomaly detection systems (\cite{yang2023prographer,han2020unicorn,manzoor2016fast}) only report anomaly results at the subgraph level, which also causes security analysts to spend a lot of time and effort confirming abnormal processes during actual deployment. In addition, Goyal et al. \cite{goyal2023sometimes} point out that this type of detection system is at risk of receiving mimicry attacks, causing attack escape (\textbf{Challenge 2}). Although the node-level detection \cite{wang2022threatrace,cheng2024kairos} achieved by the latest studies can resist mimicry attacks to a certain extent, analysts still need to spend time restoring the attack chain to filter out false positives when the false positive rate is high (\textbf{Challenge 3}). 


In recent years, large language models (LLMs) have been widely used in the field of cybersecurity, bringing a new research paradigm to the field. They have been widely applied to tasks such as threat intelligence analysis, log parsing, command line understanding, and automated response \cite{zhang2025llms,quinn2024applying,motlagh2024large,hasanov2024application}. Therefore, to address the above challenges, we propose \sn{}, an intrusion detection framework that integrates the reasoning capabilities of LLMs with the ability to model security knowledge, to achieve more accurate threat perception. First, \sn{} uses lossless compression technology to remove redundant edge information from the provenance graph while retaining semantic information. Second, \sn{} constructs a benign behavior knowledge base from benign logs and a threat intelligence knowledge base by extracting attack knowledge from cyber threat intelligence (CTI). Third, \sn{} filters out abnormal nodes and their events by constructing abnormal nodes and rare paths. \sn{} can combine the two external knowledge bases using Retrieval Augmented Generation (RAG) to enable the LLM to assess abnormal nodes, achieving node-level detection while confirming malicious nodes (\textbf{for Challenges 1, 2}). Finally, \sn{}, based on traceability mechanisms and attack graph reconstruction techniques, accurately locates the key paths and related nodes of the attack, helping analysts understand the attack process and assist in subsequent responses (\textbf{for Challenge 3}). 


The contributions of this paper are summarized as follows:
\begin{itemize}
\item  We propose a knowledge-driven provenance-based intrusion detection method.
\item  We have implemented the practice of applying LLMs to intrusion detection, emphasizing the key role of knowledge fusion in threat detection. To this end, we introduce the Retrieval-Augmented Behavior Prompting method to enhance the reasoning and judgment capabilities of LLMs.

\item  We propose \sn{}\footnote{https://github.com/Shike-Cheng/PIDS-with-LLM}, a provenance-based intrusion detection system that incorporates LLM. \sn{} is capable of node-level anomaly detection and automated reconstruction of attack graphs. Experimental results show that \sn{} can achieve performance superior to existing methods on public benchmark datasets.
\end{itemize}

\section{Background and Motivation}\label{chapter2}
\subsection {Provenance Graph}\label{sec: 2.1}
A provenance graph is a graphical structure used to record the causal relationships between events in a system \cite{li2021threat}. It can meticulously trace the chronological order of events and their invocation relations. It is widely used in forensic analysis, attack path tracking, and other tasks. A provenance graph typically relies on kernel-level system call logs. By monitoring the behavior generated by the system during runtime, it can reconstruct the interaction history between entities in the system (such as processes, files, sockets, etc.). In a provenance graph, nodes represent system objects, and edges represent causal dependencies triggered by system calls. Taking Linux as an example, kernel modules such as Auditd \cite{daniels2000network} and CamFlow \cite{pasquier2017practical} can be used to capture system call information in real time. This includes system calls like fork/clone (process creation), open/read/write (file operations), execve (program execution), connect/send/recv (network communication), and chmod/chown/setuid (permission operations). Based on these raw logs, a provenance graph can construct a complete path of system behavior, effectively detecting abnormal system behavior.As shown in Figure \ref{fig:1}, it is a typical example of a source graph, illustrating the causal interaction relationships between processes, files, and sockets in the system.


\subsection {Large Language Model}
Large Language Models (LLMs) are natural language processing technologies based on deep learning models \cite{naveed2023comprehensive}, typically employing a transformer architecture and trained on large-scale specialized corpora. By learning the statistical patterns and semantic relationships in language, LLMs demonstrate strong contextual reasoning and inference capabilities, which enable them to handle complex tasks such as question answering, text summarization, and dialogue generation.

In the field of cybersecurity, these capabilities make LLMs an emerging tool for understanding attack intentions, parsing security events, and aiding decision-making. For example, AutoAttacker \cite{xu2024autoattacker} is an LLM-guided automated network attack system that can automatically generate attack strategies and commands based on target descriptions, achieving end-to-end automation of the attack process. RACONTEUR \cite{deng2024raconteur} developed an LLM-driven shell command interpreter that can parse and semantically interpret shell commands (especially malicious ones) in combination with contextual knowledge. UniLog \cite{xu2024unilog} is an automatic logging framework based on LLMs and a context learning paradigm, which can automatically insert high-quality log statements according to program context, thereby enhancing system observability and problem tracking efficiency. LogParser-LLM \cite{zhong2024logparser} combines semantic understanding with statistical details to parse logs online without the need for hyperparameter tuning and labeled training data, demonstrating good cross-system adaptability. LLM-CloudSec \cite{cao2024llm} uses a retrieval-augmented generation mechanism and the Common Weakness Enumeration as an external knowledge base to achieve unsupervised fine-grained vulnerability analysis. Therefore, given the capabilities of LLMs in semantic reasoning and knowledge integration, we further explore their potential applications in provenance-based intrusion detection systems.


\subsection{Motivating Example}
\begin{figure*}[t]
    \centering
    \includegraphics[width=\textwidth]{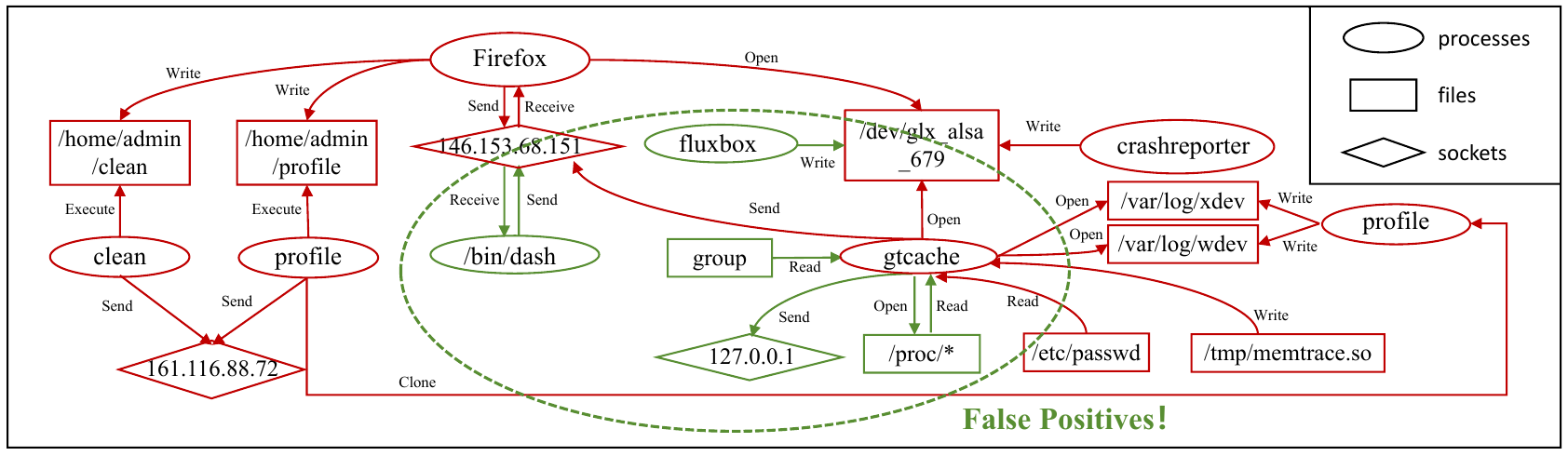}
    \caption{The DARPA E3-THEIA provenance summary graph, which provides a detailed description of the key activities in the attack example. Different shapes are used in the graph to represent different system entities: ellipses represent processes, rectangles represent files, and diamonds represent sockets. Additionally, different types of operations are labeled on the edges, including read, write, open, send, receive, clone, and execute. We have introduced different colors to represent different key information: red nodes and edges indicate confirmed attack behaviors, while green indicates identified suspicious activities, which are prone to false positives.}
    \label{fig:1}
\end{figure*}

We present a typical attack example (Figure \ref{fig:1}). The attacker exploits a remote code execution vulnerability in the Firefox browser to gain initial control over the victim's workstation and writes two malicious payloads to the local directories \verb|/home/admin/clean| and \verb|/home/admin/profile|. Subsequently, leveraging a local privilege escalation vulnerability, the attacker further obtains the system's root privileges. After acquiring elevated permissions, the attacker uses the malicious file to establish communication with the remote command and control (C\&C) server at \verb|146.153.68.151|, downloading and executing additional malicious components. The attacker also opens and writes to the device file \verb|/dev/glx_alsa_679|, utilizing this device as a covert channel for communication or data exfiltration to bypass conventional security detection.

\textbf{Limitations of Existing Techniques:} Rule-based systems \cite{milajerdi2019holmes,hossain2017sleuth} can achieve fine-grained node-level detection, but their effectiveness relies on the completeness and update frequency of the rule repository. Since rule sets often fail to cover behavioral characteristics in the system dynamically, this results in significant detection blind spots in practice and consequently low node-level accuracy. The graph-level accuracy currently used for threat hunting largely depends on the effectiveness of the graph alignment algorithm, and few works \cite{milajerdi2019poirot,altinisik2023provg} have fully open-sourced it. Learning-based anomaly detection models have been widely used by researchers to achieve graph-level or node-level detection without requiring any expert experience. However, graph-level detection \cite{han2020unicorn,manzoor2016fast} not only fails to help analysts locate threats but is also vulnerable to mimicry attacks. Moreover, anomaly detection models \cite{yang2023prographer,kairosatt,wang2022threatrace,rehman2024flash} are susceptible to concept drift, which makes it difficult for the model to distinguish between changes in benign behavior and actual malicious behavior. 

It is worth noting that current anomaly-based detection systems tend to misidentify the activities represented by the green nodes in Figure  \ref{fig:1} as threat alerts. For example, the system detects data being written to \verb|/dev/glx_alsa_679| by \verb|fluxbox|, as well as communication between \verb|/bin/dash| and the remote C\&C server. In addition, the system observes that some processes, which share the same parent process as \verb|profile|, access similar paths. Although these processes do not participate in the attack, they exhibit access patterns in the graph similar to the malicious process flow and are thus flagged by the system as potential threat nodes. This indicates that when the detection system is trained solely on benign samples, it lacks the ability to distinguish attack behaviors. Faced with overlapping behaviors among processes in complex environments, it tends to misidentify some normal activities as attack behaviors, leading to false positives.

During the application process, it is necessary to continuously retrain the model to adapt to anomalies in real business scenarios. Current technologies still face many challenges in accurately identifying complex threat behaviors, and knowledge fusion remains a trend in the latest research. To this end, \sn{} leverages the knowledge integration and reasoning capabilities of LLMs to fuse multi-source data such as system audit logs and threat intelligence, and to provide a unified representation of threat behaviors. For nodes that are prone to being misjudged as threats, such as the writing behavior of fluxbox in Figure \ref{fig:1}, the system compares and analyzes them in combination with the attack context and similar behaviors in the malicious knowledge base, further identifying that these behaviors are semantically closer to normal system activities, thereby avoiding false positives. \sn{} not only reduces the dependence on rules and manual intervention but also effectively alleviates issues such as false positives and concept drift in existing methods while accurately identifying threat behaviors.


\section{Threat Model}\label{chapter3}

We assume that the attacker has stealthiness and can use a variety of means to evade traditional detection mechanisms, including exploiting zero-day vulnerabilities and encrypting obfuscated communications. These attack methods make malicious behaviors indistinguishable from normal behaviors on the surface, making it difficult to differentiate between the two. At the same time, attackers often proceed with the attack process in multiple stages and implement it slowly over a long period of time, presenting the typical characteristic of “low-and-slow”. This means that attack behaviors are distributed over different time periods, and the overall attack chain is easily fragmented and hidden among a large number of normal behaviors, increasing the difficulty of identifying abnormal behaviors and restoring attack intentions.

In the design of \sn{}, we focus on combining system audit logs with external threat intelligence clues to form a multi-level, multi-dimensional security knowledge base. The system introduces the reasoning ability of large language models to identify malicious activities that are different from benign behaviors. To ensure the accuracy and reliability of the provenance graph construction, we also particularly emphasize the integrity and security of audit logs to prevent them from being tampered with and provide reliable support for subsequent analysis.
\afterpage{
\begin{figure*}[t]
\centering
\includegraphics[width=\linewidth]{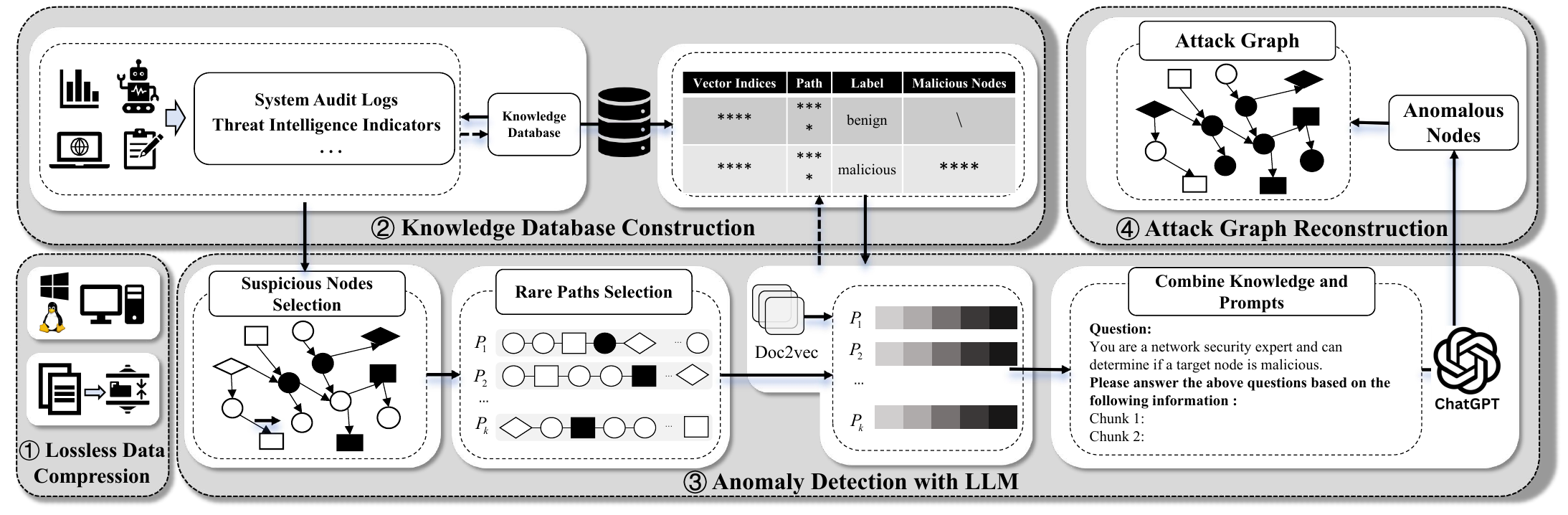}
\caption{Overview of \sn{}’ architecture.}
\label{fig:2}
\end{figure*}
}

\section{System Design}
OMNISEC is a provenance-based intrusion detection system that combines a retrieval-enhanced behavior prompting mechanism to enhance the reasoning capabilities of large language models, enabling node-level anomaly detection and supporting the automatic reconstruction of attack graphs and identification of key paths. The architecture of \sn{} is shown in Figure \ref{fig:2}, which consists of the following four core modules:

\textbf{Lossless Streaming Data Reduction (\secref{sec: DataReduction}).} Capture logs using system auditing tools, extract system entities and interaction events by parsing the logs, and merge equivalent events using data compression techniques to reduce the data processing burden while preserving the original semantics.

\textbf{Knowledge Databases Construction (\secref{sec: know_database}).} Construct an external knowledge base by combining Word2vec and Doc2vec technologies. On one hand, model path information and statistically analyze event frequencies to extract common behavioral patterns in the system. On the other hand, vectorize threat intelligence texts using natural language modeling methods to capture potential attack features. Ultimately, form a unified, structured embedded index to support subsequent behavioral matching and reasoning.

\textbf{Anomaly Detection with LLM (\secref{sec: LLM}).} Filter out suspicious nodes and rare paths using a benign path name embedding model and a rarity scoring algorithm. Further combine large language models with retrieval-augmented generation technology to semantically complete and threat-assess these candidate paths.

\textbf{Attack Graph Reconstruction (\secref{sec: traceability}).} Combine system causal structures with anomaly scoring results to filter out real attack nodes and reconstruct the complete attack graph.
\subsection{Design Goals}
The design goals of \sn{} are as follows: \textbf{G1: Lossless Streaming Data Reduction.} The system should be able to process streaming data and implement real-time data reduction. 
\textbf{G2: Knowledge Databases Construction.} Constructing knowledge databases that provide knowledge guidance for LLM. 
\textbf{G3: Anomaly Detection with LLM.} The system should be able to detect anomaly with LLM by using retrieval augmented generation (RAG) \cite{lewis2020retrieval} with the help of knowledge databases. 
\textbf{G4: Attack Graph Reconstruction.} The system should be able to reconstruct attack graphs based on anomaly detection.

\subsection{Lossless Streaming Data Reduction}\label{sec: DataReduction}
Similar to other PIDSes, \sn{} collects audit data from system audit tools (such
as event tracing for Windows \cite{ETW}, and audit for Linux platform \cite{audit}) that capture logs and construct provenance graphs that can represent topological and ordered events between entities. We focus on 3 types of entities (processes, files, and sockets) and 7 types of interaction events between entities. Among them, we convert events into edges with time and direction according to the flow of data and control flows and hold the key attribute information of the entity, as shown in Table \ref{tab:entity-edge}.
\begin{table}[h]
\centering
\caption{System entity, attributes, and dependencies.}
\label{tab:entity-edge}
\resizebox{\linewidth}{!}{%
\begin{tabular}{|c|c|c|c|}
\hline
\textbf{Subject}         & \textbf{Object} & \textbf{Edge Type} & \textbf{Entity Attributes}        \\ \hline \hline
\multirow{3}{*}{Process} & File            & Write, Execute     & \multirow{3}{*}{Process pathname, Uuid} \\ \cline{2-3}
                         & Process         & Fork               &                                   \\ \cline{2-3}
                         & Socket          & Send               &                                   \\ \hline \hline
File                     & Process         & Read, Mmap         & File pathname, Uuid                     \\ \hline \hline
Socket                   & Process         & Receive            & Source IP, Destination IP, Uuid         \\ \hline
\end{tabular}
}
\end{table}

\begin{figure}[h]
\centering
\includegraphics[width=\linewidth]{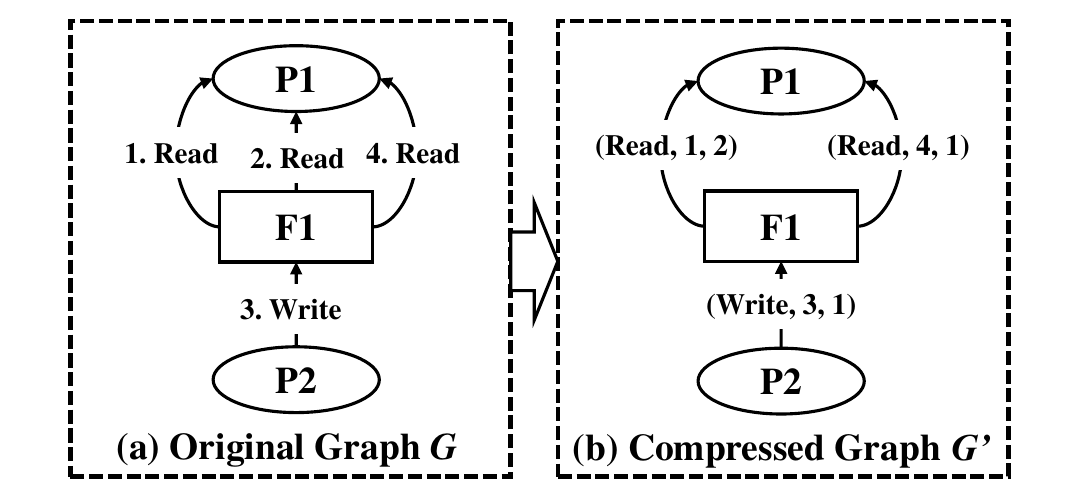}
\caption{Global semantic based compression algorithms}
\label{fig:compress_example}
\end{figure}

\par
The log parsing module sends system entities and events strictly according to the timestamp to our data reduction module as designed in Table \ref{tab:entity-edge}. The idea of data reduction that maintains global semantics from existing work \cite{zhu2021general} is borrowed in this module: only the first event that has its effect on the target vertex is preserved, provided that the semantics of the source vertex are not changed. 

It is important to emphasize that our work focuses on the change of semantic information by the temporal sequence of data flow, i.e., \sn{} focus on the event types \textit{Write}, \textit{Read}, \textit{Mmap}, \textit{Send}, \textit{Receive}. In addition, for equivalent events, \sn{} remove them as redundant events but record their event frequency as an edge attribute to be saved in the retained edges. \sn{} defines the compressed edge as $(e_t, t, c)$, where $c$ records the frequency of the event. As shown in Figure \ref{fig:compress_example}, the reduction strategy from the original graph $G$ to the compressed graph $G'$ is as follows: 1) $F1$ does not undergo a semantic change between $t=1$ and $t=2$, so the two \textit{Read} events belong to the equivalent semantics, and the edge with $t=1$ is retained and the frequency is recorded. 2) However, after $t=2$, $F1$ is written with relevant data by $P 2$, and the semantics change. Therefore, the semantics of $P1$ \textit{Read} $F1$ at $t=4$ is not equivalent to the first two \textit{Read} events and is retained.


\subsection{Knowledge Databases Construction} \label{sec: know_database}

In order to realize the best practices of anomaly detection based on knowledge fusion, our work leverages the ability of LLM to integrate knowledge. However, although LLM has demonstrated its unique advantages and broad application prospects in several fields, it is prone to produce some inaccurate information when dealing with domain-specific or highly specialized queries. Therefore, many researchers have proposed various types of methods to alleviate this problem. Among them, RAG is a technique proposed by Lewis et al. \cite{lewis2020retrieval} to effectively address this limitation. It integrates knowledge retrieved from an external knowledge base into the generation process, thus enhancing the model's ability to provide accurate and relevant responses. In this subsection we present the construction process of two external knowledge bases as follows. 
\par
\subsubsection{Benign Behavior Knowledge Base} \label{sec: benign}
In the absence of attacks, hosts within an organization generate millions of log data every day. When an attack occurs, the attacker will generate malicious operations inside the host that are different from benign application behaviors due to their special purpose. Therefore, for different application environments and business processes of enterprises, \sn{} can build a customized real-time updated benign behavior knowledge base, which can effectively reduce the misjudgment of normal behavior, thus reducing false alarms and improving the accuracy of security protection. We mainly focus on two pieces of information: 1) the path name of the underlying entity, i.e., we believe that the path information can be used as an important feature to help us characterize the behavior of the application. Therefore, \sn{} considers the path name as a sequence of text and the parts in the path (e.g., folder, file name) as words, and uses Word2vec to get the embedding vector of each node and save an embedding model. 2) The frequency of interactions of different events between the entities, which can help us to analyze the execution flow and behavioral patterns of the application through the interactions between different entities, i.e., by keeping track of the files and resources that the application accesses, they can understand the sequence of its normal operations. The total frequency of pair of nodes $(src_{name},dest_{name})$ is defined as $TotalFreq(s_n,d_n,e_t) = \sum_{t} count_{(s_n,d_n,e_t, t)}$, where $count_{(s_n,d_n,e_t, t)}$ is the frequency of $e_t$ at $t$. In addition, the total out-degree and in-degree of each node $v$ are defined respectively as:
\begin{equation}
\begin{aligned}
OUT(v) &= \sum_{d,e} TotalFreq(v_n, d, e) \\
IN(v) &= \sum_{s,e} TotalFreq(s, v_n, e)
\end{aligned}
\end{equation}
\sn{} stores the above data information into a relational database so that it can be quickly queried as a benign behavior knowledge base in subsequent attack detection.


\par
\subsubsection{Threat Intelligence Knowledge Base.} \label{sec: malicious}
Threat events recorded in CTI reports include malicious IOCs and event interactions between them. The wealth of information (tools, resources, and attack behaviors) from these attackers can provide a knowledge complement to LLM. Currently, several studies \cite{li2021attackg, satvat2021extractor, cheng2024crucialg} have parsed CTI reports. They work on constructing attack graphs similar to the structure of provenance graphs from natural language. Among them, \sn{} draws on CRUcialG \cite{cheng2024crucialg} and Extractor \cite{satvat2021extractor} to process CTI reports. \sn{} uses natural language models to construct attack scene graphs (ASGs) by extracting the attack entities and the relations between the entities from CTI reports. Therefore, \sn{} treats the longest path $path = \{v_1,e_1,v_2,e_2,...,e_{n-1},v_n\}$ composed of nodes and edges in ASGs as a sentence and embeds the path using the Doc2vec. In summary, \sn{} converts malicious attack events into malicious paths and projects them into a numeric vector space to be stored in a database constituting a threat intelligence knowledge base.

\subsection{Anomaly Detection with LLM} \label{sec: LLM}
In the previous modules, we have constructed external knowledge bases for benign (\secref{sec: benign}) and malicious (\secref{sec: malicious}) knowledge, respectively, in this subsection we will describe the LLM based anomaly detection process of \sn{} using RAG.
\par
\subsubsection{Selection of Suspicious Nodes} 
The provenance graph grows in the number of entities over time, and to narrow down the scope of anomalies, \sn{} screens the suspicious nodes in the provenance graph. Usually, processes often load or access the same set of libraries and configuration files at startup, or generate some temporary files and pipes during execution. These objects become noise information for suspicious node filtering. Therefore, \sn{} uses the benign path name embedding model to embed the path names of the nodes in the monitoring process and calculate the cosine similarity with the nodes in the benign behavior database to filter out the suspicious nodes (suspicious nodes filtered out are often those that have not appeared in the knowledge base).


\subsubsection{Selection of Rare Paths} 
The contextual information of the control and data flows interacting between nodes determines whether the node is involved in malicious behavior or not. However, the number of paths is exponentially related to the number of nodes, and directly extracting all the paths of anomalous nodes from the provenance graph may lead to the dependency explosion problem. To solve this problem, a path selection algorithm based on rarity scoring is proposed. For event $e_i=(v_{s}, v_{d}, e_t)$, the frequency score of $e_i$ is defined as: 
\begin{equation}\label{equ: num2}
\begin{aligned}
S(e_i) = D_{out}(v_s)F(e_i)D_{in}(v_d)
\end{aligned}
\end{equation}
In Equation \ref{equ: num2}, $F(e_i)$ is defined as:
\begin{equation}\label{equ: num3}
\begin{aligned}
F(e_i)=\frac{TotalFreq(s_n, d_n, e_t)}{\sum_{d, e} TotalFreq(s_n, d, e)}
\end{aligned}
\end{equation}
$s_n$ and $d_n$ represent the path names of the subject and object nodes, respectively. $F(e_i)$ is refers to the percentage of $e_i$ in all events of $s_n$ in the host. $D_{out}(v_s)$ and $D_{in}(v_d)$ are defined respectively as:
\begin{equation}\label{equ: num4}
\begin{aligned}
D_{out}(v_s) = \frac{OUT(s_n)}{\sum_{i} OUT(v_i)}, D_{in}(v_d) = \frac{IN(d_n)}{\sum_{i} IN(v_i)}
\end{aligned}
\end{equation}
In order to reduce the complexity of path search and to ensure real-time detection, each anomalous node searches for $k_1$-hop neighbors forward and backward, respectively, according to the time sequence. Thus, for any anomalous node $v$ with path $p_v=\{e_1,e_2, ..., e_{k_1},..., e_{2k_1}\}$, the rarity score is defined as:
\begin{equation}\label{equ: num5}
\begin{aligned}
R(p_v) = -\log_{2}{\prod_{i=1}^{2k_1}S(e_i)}
\end{aligned}
\end{equation}
By defining the rarity score, \sn{} selects the top $k_2$ paths with the highest anomaly scores among the paths of the suspect nodes in the order of events. We also pay extra attention to these non-suspicious nodes involved in these rare paths. Similarly, \sn{} will look for high-frequency (lowest anomaly scores) events where these nodes are involved and form benign paths in the benign behavior database. The benign paths will also be encoded using the Doc2vec model, and then together with the threat intelligence knowledge base to form a vector-indexed corpus for the RAG.

\par
\subsubsection{RAG with LLM} 
The LLM represented by ChatGPT proposed by OpenAI \cite{openai} is gradually used by various small domains to fine-tune it for downstream tasks. However, with the introduction of GPT-4, the cost of fine-tuning the large model gradually rises. Therefore, we draw on the typical RAG application workflow proposed by Gao et al. \cite{gao2023retrieval} to optimize the node-level anomaly detection to enhance the judgment capability of the LLM via the RAG as shown in Figure \ref{fig:2}. First of all, the suspect node and its rare path will be entered into the LLM as a user query. The RAG will use the same encoding as in the indexing phase, i.e., it will transcode the paths of the query using Doc2vec. Next, it will proceed to compute the similarity scores between the query vectors and the vectorized blocks in the indexed corpus. \sn{} prioritizes and retrieves the top $K$ blocks that are most similar to the query. Finally, \sn{} synthesizes the query with the retrieved relevant benign and malicious knowledge into a coherent cue to be fed into the LLM for anomalous node determination (prompt template and its structure can be found in Figure \ref{fig:2}).

\subsection{Attack Graph Reconstruction} \label{sec: traceability}
Attack reconstruction has proved useful in reducing the time for analysts to perform forensic analysis as well as in understanding the complete attack story. \sn{} performs anomaly judgment on suspicious nodes with LLM. However, anomalies are not equal to real attacks. \sn{} uses causal analysis methods (we draw on some label-based PIDSes \cite{milajerdi2019holmes,zhu2023aptshield} for representation of attack stages) to filter out the real attack nodes and reconstruct the complete attack graph. The key insight of \sn{} in this process is that the attacker's complete attack activity relies on the contextualization. Therefore, the real attack nodes must have key events associated with each other (i.e., the real malicious nodes are connected to each other relying on the attack-related nodes, and we treat these attack-related nodes as an important part of the attack reconstruction as well). As shown in lines 1 to 25 in Algorithm \ref{alg:ag_cons}, \sn{} processes the LLM detection results, finds anomalous clusters related between nodes based on rare paths and reduces the attack graph based on causal relations between suspicious flows.

\begin{algorithm}[h]
    \SetAlgoLined 
    \caption{Attack Graph Reconstruction}
    \label{alg:ag_cons}
    \KwIn{Anomalous Nodes List $V_a$\; Key\_word = ['/etc/passwd', 'hostname', ...]\; Rare Event List (REL)\; Rare Event Score (RES)\;}
    \KwOut{Attack Graph $G$;}
    \For{$v_i \in V_a$}{
	\For{$v_j \in V_a$}{
            \If{$REL_{V_i} \cap Key\_word \neq \emptyset$ and $REL_{V_j} \cap Key\_word \neq \emptyset$}{
                $v_i, v_j \rightarrow cluster$\;
            }
            \If{$v_j \in REL_{V_i}$}{
                $v_i, v_j \rightarrow cluster$\;
            }
        }
    }
    Merge related node $Clusters$\;
    \For{$c \in Clusters$}{
        \eIf{$len(c)==1$}{
            continue\;
        }{
            \For{$v \in c$}{
                \For{$p \in REL_v$}{
                    $P_{score} += RES_p$\;
                }
            }
        }
        $c_{score} = P_{score} / \text{len}(c)$\;
    }
    $Attack Nodes = \max(c_{score})$\;
    $Attack Graph Reconstruction(Attack Nodes, REL)$;
\end{algorithm}

\section{Evaluation}
In this section, we first present the evaluation preparation, including the introduction of the dataset, the acquisition of ground truth, and the setting of the evaluation metrics. Our evaluation will be organized around answering the following questions: 
\begin{itemize}
    \item \textbf{RQ1:} Can \sn{} accurately detect attacks and how does it compare to the state-of-the-art?
    \item \textbf{RQ2:} What is the size of the attack graph for \sn{} reconstruction?
    \item \textbf{RQ3:} What is \sn{}'s end-to-end overhead?
    \item \textbf{RQ4:} How do hyperparameters affect \sn{}'s performance and overhead? 
\end{itemize}

\subsection{Evaluation Preparation}
We deploy our implementation of \sn{} on a computer with Intel (R) Core (TM) i9-10900K CPU @ 3.70GHz and 64GB memory. The LLM in the \sn{} calls the “gpt-3.5-turbo” API from OpenAI ChatGPT \cite{openai}. 
\par
\textbf{DARPA Dataset.} We obtain our experimental datasets from DARPA Transparent Computing \cite{DARPA-TC}. From October 2016 to May 2019, the DARPA TC program has organized five larger scale red and blue adversarial engagement exercises. Different technical teams operate around different operating systems to execute real APT attacks by simulating corporate networks. For the experimental evaluation of \sn{}, we choose the part attack data from CADETS and THEIA in Engagement \#3 \cite{DARPA-TC-E3} and \#5 \cite{DARPA-TC-E5}. To ensure the fairness of the comparison experiments, we test on the same datasets as state-of-the-art. For example, ThreaTrace \cite{wang2022threatrace} and Flash \cite{rehman2024flash} only expose part of the trained weights model in DARPA Engagement \#3.
\par
\textbf{Ground Truth.} DARPA Engagement \#3 and \#5 provide ground truth reports for researchers to label attack nodes. Currently, some studies have manually labeled ground truth to evaluate PIDS performance. For example, Kairos provides supplementary material \cite{kairosatt} and Jason et al. \cite{reapr-ground-truth} publish standardizing ground truth labels. In the evaluation of \sn{}, we refer to these studies and open-source our ground truth (we publish the uuid of the manually labeled attack nodes)\footnote{https://github.com/Shike-Cheng/PIDS-with-LLM/tree/main/Ground\_Truth}. In contrast to these studies, we label all attack nodes from campaigns implemented by each team (i.e., CADETS and THEIA) at different times (i.e., these teams implemented various attacks at different times, and some open-source PIDSes \cite{wang2022threatrace,rehman2024flash} also chose different evaluation time periods on the same dataset). Table \ref{tab:attack_datasets} details the 6 attack scenarios we selected from DARPA and reports the number of attack nodes we labeled.

\begin{table}[]
\centering
\caption{Detail of attack scenarios in DARPA datastes.}
\label{tab:attack_datasets}
\resizebox{\linewidth}{!}{%
\begin{tabular}{|c|c|c|c|}
\hline
\textbf{Datasets} & \textbf{Description}                                                                             & \textbf{\begin{tabular}[c]{@{}c@{}}Timestamp Fragment \\ (yyyy-mm-dd HH:MM:SS)\end{tabular}} & \textbf{\begin{tabular}[c]{@{}c@{}}\# of Attack\\ Nodes\end{tabular}} \\ \hline \hline
E3-CADETS-1       & \multirow{2}{*}{\begin{tabular}[c]{@{}c@{}}Nginx Backdoor with Drakon \\ In-Memory\end{tabular}} & \begin{tabular}[c]{@{}c@{}}2018-04-06 00:00:00 -\\ 2018-04-07 00:00:00\end{tabular}          & 75                                                                   \\ \cline{1-1} \cline{3-4} 
E3-CADETS-2       &                                                                                                  & \begin{tabular}[c]{@{}c@{}}2018-04-11 16:36:27 -\\ 2018-04-12 22:24:07\end{tabular}          & 80                                                                   \\ \hline \hline
E3-THEIA-1        & \begin{tabular}[c]{@{}c@{}}Firefox backdoor with drakon\\ APT in memory\end{tabular}             & \begin{tabular}[c]{@{}c@{}}2018-04-10 00:00:00 -\\ 2018-04-11 00:00:00\end{tabular}          & 57                                                                   \\ \hline
E3-THEIA-2        & \begin{tabular}[c]{@{}c@{}}Browser extension with drakon APT\\ on disk\end{tabular}              & \begin{tabular}[c]{@{}c@{}}2018-04-12 10:35:21 -\\ 2018-04-12 14:57:12\end{tabular}          & 65                                                                   \\ \hline \hline
E5-CADETS         & Firefox Drakon APT (Failed)                                                                      & \begin{tabular}[c]{@{}c@{}}2019-05-17 00:00:00 -\\ 2019-05-17 11:00:00\end{tabular}          & 11                                                                   \\ \hline
E5-THEIA          & \begin{tabular}[c]{@{}c@{}}Firefox Drakon APT BinFmt-Elevate\\ Inject\end{tabular}               & \begin{tabular}[c]{@{}c@{}}2019-05-15 00:00:00 -\\ 2019-05-16 00:00:00\end{tabular}          & 33                                                                   \\ \hline
\end{tabular}%
}
\end{table}

\par
\textbf{Evaluation Metrics.} Since anomalous subgraphs or time windows still require further analysis if the analyst wants to locate the real attack event, we believe that node-level evaluation granularity better reflects the detection performance of PIDSes. We measure \textit{true positives} (TPs), \textit{false positives} (FPs), \textit{true negatives} (TNs) and \textit{false negatives} (FNs). TP refers to nodes that are recognized by \sn{} as true attacks, FP refers to nodes that \sn{} recognizes as attack but are not, TN refers to nodes that \sn{} correctly recognizes as non-attacks, and FN refers to nodes that \sn{} recognizes as non-attacks but are real attacks. 

\begin{table}[]
\centering
\caption{Experimental evaluation results of \sn{} at the node-level.}
\label{tab:detect_data}
\resizebox{\linewidth}{!}{%
\begin{tabular}{|c|c|c|c|c|c|c|c|c|}
\hline
\textbf{Datasets} & \textbf{TP} & \textbf{TN} & \textbf{FP} & \textbf{FN} & \textbf{Precision} & \textbf{Recall} & \textbf{Accuracy} & \textbf{F1-score} \\ \hline \hline
E3-CADETS-1       & 74          & 55956       & 0           & 1           & 1.000              & 0.987           & 1.000             & 0.993             \\ \hline
E3-CADETS-2       & 71          & 55950       & 2           & 9           & 0.973              & 0.888           & 1.000             & 0.928             \\ \hline \hline
E3-THEIA-1        & 56          & 112223      & 0           & 1           & 1.000              & 0.982           & 1.000             & 0.991             \\ \hline
E3-THEIA-2        & 60          & 53982       & 0           & 5           & 1.000              & 0.923           & 1.000             & 0.960             \\ \hline \hline
E5-CADETS         & 8           & 528,201     & 0           & 3           & 1.000              & 0.727           & 1.000             & 0.842             \\ \hline
E5-THEIA          & 31          & 157392      & 13          & 2           & 0.705              & 0.939           & 1.000             & 0.805             \\ \hline
\end{tabular}%
}
\end{table}

\subsection{Detection Performance}\label{sec: 5.2}
To respond \textbf{RQ1}, we simulate \sn{} to monitor the underlying anomalous behavior of the host in real time and evaluate \sn{} performance on each dataset. In addition, we selected the SOTA PIDSes Kairos \cite{10646673}, ThreaTrace \cite{wang2022threatrace}, and Flash \cite{rehman2024flash} for comparison with \sn{}. The reasons for choosing these studies are as follows: 1) These studies have released their high-quality and complete open-source code for us to reproduce locally. 2) They achieve a fine-grained level of anomalous detection. 3) As Kairos said, other work is closed source or missing key component code. 
\par
Table \ref{tab:detect_data} shows the precision, recall, accuracy, and F1-score results for all datasets. We compute these metrics based on node level. From Table \ref{tab:detect_data}, we can see that \sn{} is able to accurately detect most of the attacks and produce fewer false positives. In E5-THEIA, \sn{} has more obvious FPs. Through our analysis, 
these FPs are nodes affected by the malicious process “sshd”. That is, when the malicious process “sshd” is labeled as a malicious node by LLM, the node involved in its subsequent rare events is also reconstructed as part of the attack. However, even though the inclusion of FPs affects the accuracy, it does not prevent security analysts from analyzing the reconstructed attack graph.
\par

\begin{table}[]
\centering
\caption{Comparison with the Kairos in terms of window-level detection accuracy.}
\label{tab:comparsion_kairos}
\resizebox{\linewidth}{!}{%
\begin{tabular}{|c|c|c|c|c|c|c|}
\hline
\textbf{Datasets}            & \textbf{Systems} & \textbf{Precision} & \textbf{Recall} & \textbf{F1-score} & \textbf{\# of Alert Node} & \textbf{\% Alert Node} \\ \hline \hline
\multirow{2}{*}{E3-CADETS-1} & \sn{}             & 1.00               & 1.00            & 1.00              & 74                       & 0.1321                \\ \cline{2-7} 
                             & Kairos          & 0.80               & 1.00            & 0.89              & 221                      & 0.3944                \\ \hline \hline
\multirow{2}{*}{E3-THEIA-1}  & \sn{}             & 1.00               & 1.00            & 1.00              & 56                       & 0.0499                \\ \cline{2-7} 
                             & Kairos          & 0.80               & 1.00            & 0.89              & 5,525                    & 4.9207                \\ \hline \hline
\multirow{2}{*}{E5-CADETS}   & \sn{}             & 1.00               & 1.00            & 1.00              & 8                        & 0.0015                \\ \cline{2-7} 
                             & Kairos          & 1.00               & 1.00            & 1.00              & 660                      & 0.1249                \\ \hline \hline
\multirow{2}{*}{E5-THEIA}    & \sn{}             & 0.50               & 1.00            & 0.67              & 44                       & 0.0279                \\ \cline{2-7} 
                             & Kairos          & 0.67               & 1.00            & 0.80              & 382                      & 0.2426                \\ \hline
\end{tabular}%
}
\vspace{-10pt}
\end{table}

\textbf{Comparison Study.} Comparing systems fairly with SOTA is difficult due to differences in evaluation protocols. Therefore, we respect the protocols of each system's evaluation and compare the performance of the systems as fairly as possible. Table \ref{tab:comparsion_kairos} and Table \ref{tab:comparsion} show the performance of the \sn{} against Kairos, ThreaTrace and Flash on the Darpa dataset.
\par
\textbf{Kairos.} \sn{} uses time windows to compute the evaluation metrics (the generated attack graphs are divided according to time windows), as in Kairos (the authors consider the time window containing the manually labeled ground truth to be anomalous). During the attack reconstruction process, Kairos uses anomalous triples whose reconstruction error exceeds the threshold for performing community discovery. Since Kairos still needs expert knowledge to find the real attack after generating the candidate summary graph, we report the number of anomalous alert nodes in the table. As can be seen from Table \ref{tab:comparsion_kairos}, excluding E5-THEIA, \sn{} exhibits comparable or even better performance than Kairos. The reason why \sn{} exhibits higher FPs in E5-THEIA is as we analyze earlier. Nevertheless, the number of alert nodes reported by \sn{} is much smaller than that of Kairos, and sysadmins can directly analyze the attack graphs reconstructed by \sn{} without having to spend time sifting through candidate summary graphs. Kairos emphasizes that it does not require any prior knowledge, but we believe that relying solely on learning benign behavior patterns leads to a large number of FPs in the results. Although Kairos uses community discovery algorithms to simplify the detection results into candidate summary graphs, removing FPs requires not only manual analysis of these summary graphs by analysts but also retraining of the model with FPs.
\par

\begin{table}[]
\centering
\caption{Comparison with the ThreaTrace and Flash in terms of node-level detection accuracy.}
\label{tab:comparsion}
\resizebox{\linewidth}{!}{%
\begin{tabular}{|c|c|c|c|c|c|c|c|}
\hline
\textbf{Datasets}            & \textbf{Systems} & \textbf{Precision} & \textbf{Recall} & \textbf{F1-score} & \textbf{\begin{tabular}[c]{@{}c@{}}\# of Alert\\ Node\end{tabular}} & \textbf{\begin{tabular}[c]{@{}c@{}}\% Alert\\ Node\end{tabular}} & \textbf{\begin{tabular}[c]{@{}c@{}}\# of Attack Nodes \\ in Ground Truth\end{tabular}} \\ \hline \hline
\multirow{3}{*}{E3-CADETS-2} & \sn{}              & 0.97               & 0.89            & 0.93              & 73                                                                 & 0.0204                                                          & 80                                                                                    \\ \cline{2-8} 
                             & ThreaTrace       & 0.90               & 0.99            & 0.95              & 20,104                                                             & 5.6286                                                          & \multirow{2}{*}{12,858}                                                               \\ \cline{2-7}
                             & Flash            & 0.95               & 0.99            & 0.97              & 13,854                                                             & 3.8788                                                          &                                                                                       \\ \hline \hline
\multirow{3}{*}{E3-THEIA-2}  & \sn{}              & 1.00               & 0.92            & 0.96              & 60                                                                 & 0.0210                                                          & 65                                                                                    \\ \cline{2-8} 
                             & ThreaTrace       & 0.87               & 0.99            & 0.93              & 28,134                                                             & 9.8518                                                          & \multirow{2}{*}{25,363}                                                               \\ \cline{2-7}
                             & Flash            & 0.93               & 0.99            & 0.96              & 27,591                                                             & 9.6616                                                          &                                                                                       \\ \hline
\end{tabular}%
}
\vspace{-10pt}
\end{table}

\textbf{ThreaTrace and Flash.} ThreaTrace \cite{wang2022threatrace} and Flash \cite{rehman2024flash} can achieve node-level detection. And Flash uses the ground truth published by ThreaTrace and its reported number of malicious nodes is comparable to that of ThreaTrace. As stated by Kairos, authors of ThreaTrace manually mark nodes in the ground truth and their 2-hop ancestors/descendants as anomalous, even if neighboring nodes are not involved in the attack. In the ground truth published by ThreaTrace, there are 12,858 malicious nodes in CADETS and 25,363 malicious nodes in THEIA (more than 100 times the number of nodes we labeled). In particular, 22,250 malicious nodes we find reported by ThreaTrace in E3-THEIA-2 are isolated nodes that \sn{} is unable to locate without event interactions. In addition, Wang et al. \cite{wang2024incorporating} also point out that ThreaTrace and Flash do not specify a data labeling policy or context, and that the number of malicious labels tagged is excessive (e.g., more than 12,000 nodes are tagged as malicious during the 30 hours of CADETS), which is impractical for a real-world SOC. Therefore, to ensure fairness, we do not evaluate these three systems on any evaluation protocol. Instead, we care more about how much the detection results can help analysts reduce manual labour costs. As shown in Table \ref{tab:comparsion}, the number of alert nodes in \sn{} can be far less than ThreaTrace and Flash while ensuring low FNs. In our opinion, the time it takes for sysadmins to find FNs by tracing back the known attack graph is much less than the time it takes to sift out FPs from more than 15K alert nodes on average.

\subsection{Reconstruction Performance}
The reconstruction of the attack graph not only provides PIDS with the explanation of the attack but also helps security analysts quickly grasp the whole picture of the attack. \sn{} is able to find all malicious nodes that constitute an attack, and generates the attack graph based on the judgment result of LLM combined with the traceability analysis automatically. In particular, we take all event interactions between malicious nodes as the edges of the attack graph. This is because even if the system generates false alarm nodes, the analyst can rule them out by observing interaction events between the false alarm nodes and real attack nodes. To respond \textbf{RQ2}, as shown in Table \ref{tab:graph_size}, compared to the original provenance graph, we generate the attack graph to be small (where the edge parsimony reaches 67K times). 

\begin{table}[]
\centering
\caption{Size of attack graphs on all DARPA dataset.}
\label{tab:graph_size}
\resizebox{\linewidth}{!}{%
\begin{tabular}{|c|c|c|c|c|}
\hline
\textbf{Dataset} & \textbf{\# of Attack Nodes} & \textbf{\# of Attack Edges} & \textbf{\# of Total Nodes} & \textbf{\# of Total Edges} \\ \hline \hline
E3-CADETS-1      & 74                         & 246                        & 56,031                     & 311,587                    \\ \hline 
E3-CADETS-2      & 73                         & 295                        & 56,032                     & 323,192                    \\ \hline \hline
E3-THEIA-1       & 56                         & 108                        & 112,280                    & 662,882                    \\ \hline
E3-THEIA-2       & 60                         & 117                        & 54,047                     & 269,047                    \\ \hline \hline
E5-CADETS        & 8                          & 10                         & 528,212                    & 3,897,079                   \\ \hline \hline
E5-THEIA         & 44                         & 504                        & 157,438                    & 2,528,278                   \\ \hline
\end{tabular}%
}
\end{table}

\subsection{End-to-end Overhead} \label{eval:overhead}
In this section, we focus on the overhead of \sn{} to response \textbf{RQ3}. The overhead of \sn{} is mainly caused by the following aspects: 1) the storage overhead of the application behavior database and the threat intelligence database; 2) the time overhead of \sn{} to construct rare paths for the suspicious nodes during attack detection; and 3) the capital overhead of \sn{} to call gpt-3.5-turbo's API for suspicious nodes judgement.
\par
\textbf{Storage Overhead.} As described in Section \ref{sec: know_database}, only the node characteristics from the benign logs and the interaction characteristics between the nodes are stored in the application behavior Database, including the embedding vectors of the node path names, the frequency of interactions between the nodes, and the out-degree and in-degree of the nodes. Table \ref{tab:database_size} shows the benign days segments selected by \sn{} on each dataset, the size of the application behavior database, and the size of the database in terms of the original dataset, respectively. As shown in Table \ref{tab:database_size}, the application behavior knowledge base occupies only a tiny portion of the original dataset. This helps the system to significantly reduce the storage overhead. In addition, the threat intelligence database stores the malicious paths as well as path embedding vectors for constructing the attack scene graphs from CTI reports. Currently, we have collected 5,231 CTI reports and constructed 10,532 malicious paths from them, and the database size is 45.27M.
\par

\begin{table}[]
\centering
\caption{Size of application behavior database on all DARPA datasets. The fourth column shows the size of the database as a percentage of the original dataset.}
\label{tab:database_size}
\resizebox{\linewidth}{!}{%
\begin{tabular}{|c|c|cc|}
\hline
\multirow{2}{*}{\textbf{Dataset}} & \multirow{2}{*}{\textbf{Days (yyyy-mm-dd)}} & \multicolumn{2}{c|}{\textbf{Size of Knowledge Database}}              \\ \cline{3-4} 
                                  &                                             & \multicolumn{1}{c|}{\textbf{Database}} & \textbf{Database / Original} \\ \hline \hline
E3-CADETS                         & 2018/04/02-2018/04/04                       & \multicolumn{1}{c|}{24.58MB}             & 0.37\%                        \\ \hline \hline
E3-THEIA                          & 2018/04/03-2018/04/05                       & \multicolumn{1}{c|}{33.92MB}             & 0.12\%                        \\ \hline \hline
E5-CADETS                         & 2019/05/08-2019/05/11                       & \multicolumn{1}{c|}{250.75MB}            & 0.09\%                        \\ \hline \hline
E5-THEIA                          & 2019/05/08-2019/05/09                       & \multicolumn{1}{c|}{20.76MB}             & 0.13\%                        \\ \hline
\end{tabular}%
}
\end{table}

\textbf{Runtime Overhead.} \sn{} processes the streaming system log, feeds the suspicious nodes and their rare paths to the LLM for judgment. Meanwhile, the main time overhead affecting \sn{} for real-time detection is imposed by the selection of rare paths. We show in Table \ref{tab:time_overhead} the minimum time for \sn{} to construct a rare path for a suspicious node on each dataset, the maximum time, the total time to construct all rare paths, and the average time to construct a rare path for each suspicious node. From the results in Table \ref{tab:time_overhead}, we can find that the time for rare path construction is not very much correlated with the number of rare nodes. Upon analysis, we believe that it depends on the complexity of the events in which the suspicious nodes are involved. For example, in E5-THEIA, the types of suspicious nodes are mostly sockets and these nodes are involved in fewer events in the logs. Throughout the experiments, the maximum average time for the system to process a suspicious node is 0.71s, and the maximum time to process all anomalous nodes is 365.6s (less than the maximum time required by Kairos to process a time window, 376.2s). Therefore, we consider this time to be an acceptable time to achieve real-time detection.
\par

\begin{table}[]
\centering
\caption{Rare paths are constructed in all datasets at all times. The second column indicates the number of suspicious nodes. The third, fourth and sixth columns denote the minimum and maximum time to find a rare path for a suspicious node, respectively. The fifth column represents the total time to find a rare path for all the suspicious nodes.}
\label{tab:time_overhead}
\resizebox{\linewidth}{!}{%
\begin{tabular}{|c|c|c|c|c|c|}
\hline
\textbf{Dataset} & \textbf{\# of Suspicious Nodes} & \textbf{Min (ms)} & \textbf{Max (s)} & \textbf{SUM (s)} & \textbf{AVG (s)} \\ \hline \hline
E3-CADETS-1      & 438                            & 0.00004           & 14.33000         & 310.01427        & 0.70770          \\ \cline{1-6}
E3-CADETS-2      & 98                             & 0.00001           & 11.41969         & 54.63248         & 0.55747          \\ \hline \hline
E3-THEIA-1       & 548                            & 0.00001           & 146.75913        & 365.64470        & 0.66723          \\ \cline{1-6}
E3-THEIA-2       & 131                            & 0.00002           & 7.72443          & 34.91193         & 0.26650          \\ \hline \hline
E5-CADETS        & 9                              & 0.00003           & 0.00019          & 0.00208          & 0.00023          \\ \hline \hline
E5-THEIA         & 913                            & 0.19145           & 48.59543         & 59.88906         & 0.06560          \\ \hline
\end{tabular}%
}
\vspace{-8pt}
\end{table}

\textbf{Monetary Expenses.} To the best of our knowledge, \sn{} is the first intrusion detection system to apply LLM at the provenance graph level. \sn{} calls the API of OpenAI's gpt-3.5-turbo for suspicious node research. According to OpenAI's official pricing standards, the input pricing of gpt-3.5-turbo is \$0.5 / M, and the output pricing is \$1.5 / M (1 M = 1 million token). Upon analysis, the monetary overhead of \sn{}'s inputs is mainly brought about by the number of rare paths of the suspicious nodes (although benign and malicious paths matched from the knowledge database also affect the number of tokens, this is also mainly affected by the number of rare paths). We restrict the output of the LLM to “benign” or “malicious” in the input prompt. We count the number of rare paths in each dataset and their corresponding monetary overheads in Table \ref{tab:gpt_money}. As shown in Table \ref{tab:gpt_money}, the money overhead is proportional to the number of rare paths. In addition, we believe that users can reduce the number of tokens by continuously optimizing the prompt, which leads to less overhead.


\begin{table}[]
\centering
\caption{The number of rare paths and the total amount spent by \sn{} on all DARPA datasets.}
\label{tab:gpt_money}
\resizebox{\linewidth}{!}{%
\begin{tabular}{|c|c|c|}
\hline
\textbf{Dataset} & \textbf{\# of Rare Paths} & \textbf{US\$ Spent (gpt-3.5-turbo)} \\ \hline \hline
E3-CADETS-1      & 3,775                      & 0.150               \\ \cline{1-3}
E3-CADETS-2      & 481                       & 0.030               \\ \hline \hline
E3-THEIA-1       & 618                       & 0.100               \\ \cline{1-3}
E3-THEIA-2       & 831                       & 0.120               \\ \hline \hline
E5-CADETS        & 5                         & 0.001               \\ \hline \hline
E5-THEIA         & 55                        & 0.002               \\ \hline
\end{tabular}%
}
\vspace{-8pt}
\end{table}

\subsection{Hyperparameter Impact on Performance and Overhead} \label{eval:Hyperparameter}
\begin{figure*}[h!t]
\centering
\includegraphics[width=\linewidth]{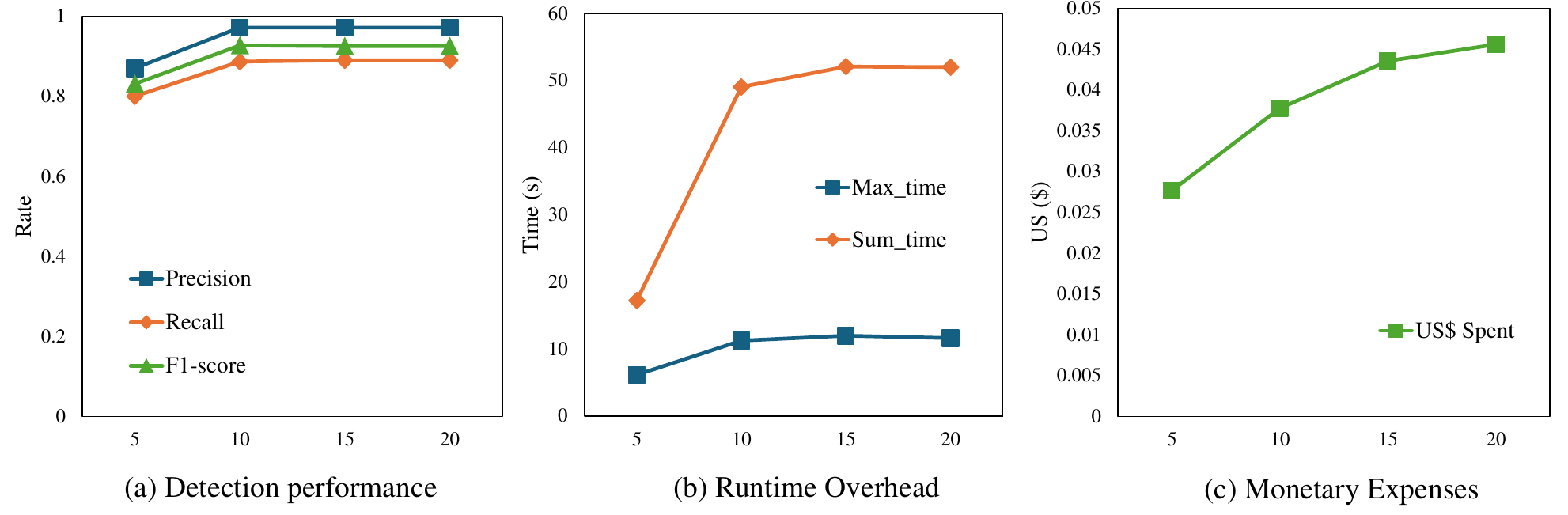}
\caption{The performance (detection performance, runtime overhead, and monetary expenses) of \sn{} on E3-CADETS-2 with varying $k_1$.}
\label{fig:tiaocan}
\vspace{-12pt}
\end{figure*}

The number of neighbor hops of a suspicious node is the key to the performance of \sn{}. To response \textbf{RQ4}, Figure \ref{fig:tiaocan}-(a) shows the effect of the number of neighbor nodes $k_1$ on the detection performance of the system on E3-CADETS-2. From the experimental results, it is clear that the detection performance improves significantly as we increase the number of neighboring nodes. However, after $k_1$ reaches 10, the performance improvement is no longer significant. This is due to the fact that most of the nodes in the dataset have fewer than 15 neighbor nodes. when $k_1$ is increased above 15 there is hardly any better performance. Both Figure \ref{fig:tiaocan}-(b) and Figure \ref{fig:tiaocan}-(c) illustrate the effect of parameter $k_1$ on the system overhead. As the number of neighbor nodes grows, the rare path finding time increases. This is because the growth of neighbor nodes increases the time complexity of the path algorithm search. Secondly, the number of neighbor nodes equally affects the length of rare path construction, which in turn increases the token spend. Finally, we find $k_1=10$ to be ideal among all datasets. Similarly, we get the optimal $k_2=10$ and $K=5$.

\section{Discussion}\label{sec:discussion}
In this work, we approach from the perspective of knowledge modeling. We explore the feasibility and methodological approaches of integrating large language models into host-based intrusion detection systems. We analyzed the challenges faced by existing PIDSes in practical applications, including the dependence of rule-based methods on predefined rules, the demand for high-quality labeled data by supervised learning methods, and the difficulty of anomaly detection methods in distinguishing between normal variations and real attack behaviors. We propose an insight: leveraging LLM to facilitate multi-source security knowledge modeling and reinforce the “last line of defense” in threat detection. We also validate this through the application of LLM in threat detection. Our practice also provides a possibility for applications of LLM in PIDSes. However, there are still some future challenges for research in this direction.
\par
\textbf{LLM Hallucination.} When a LLM is presented with a question that is not supported by sufficient facts, it may produce false or misleading outputs when answering it (i.e., the “LLM hallucination” \cite{maynez2020faithfulness}). In threat detection, hallucinations are often caused by the fact that we do not give LLM sufficient external threat knowledge (or sufficiently fine-tuned training datasets). In real threat scenarios, unfounded answers from LLM may result in a large number of false alarms. Therefore, to address such challenges, we need to enrich the knowledge database (for local models, we need to introduce diverse datasets during model training) and provide as much relevant knowledge as possible to enhance the judgment of LLM.
\par
\textbf{LLM Poisoning.} Attack escape is a strategy used by an attacker to avoid the surveillance of the detection system. Thus, an attacker may try to bypass the detection system by confusing LLM with adversarial samples. In our scenario, there are two possible ways for an attacker to poison. The first type of poisoning is when an attacker poisons a continuously updated benign behavior knowledge base by performing some low and slow action (suspicious but not triggering an alert). Thus, we need update the knowledge base in a trusted environment to avoid poisoning. Another form of poisoning is a new threat called “retrieval poisoning” \cite{zhang2024human} , targeting LLM-powered applications, which exploits the design features of LLM application frameworks to perform imperceptible attacks during RAG. These authors suggest two possible defenses and point out developing more effective defense mechanisms remains a critical need.


\textbf{LLM Overhead.} The computational and storage requirements of LLM are very high. When we use online LLM for real-time detection, we need to consider not only the network latency problem, but also the resource overhead caused by the number of tokens fed into the LLM. When we use offline LLM, we need to consider the training cost of local base LLM and the problem of insufficient resources or high latency during real-time detection. To address such issues, we can use distillation models (small models) to help us reduce the computational requirements. For example, in our practice, the selection of suspicious nodes and rare paths in the scoring system can be replaced by small models. In addition, small models can also be used to help us generate the best knowledge prompts.

\textbf{Data Privacy and Compliance.} Threat detection typically involves large amounts of sensitive data, such as user activity logs, system configurations, etc. When using LLM, data transfer and storage risks may be involved, which are especially evident in cloud-based models. Therefore, to ensure that models and data are stored and processed in compliance with relevant privacy regulations, we should try to deploy models locally or anonymize data to prevent sensitive information leakage.





\section{Related Work}\label{sec:related_work}

In this section, we summarize and analyze the existing PIDSes into two categories (heuristic-based and learning-based). Additionally, to our knowledge, \sn{} is the first to apply LLM to PIDS, and research on LLM in other security fields has also provided us with inspiration for future improvements.

\textbf{Heuristic-based Instrusion Detection.} Holmes \cite{milajerdi2019holmes} matches rules based on the information flow between low-level entities in the system according to the TTP patterns in the ATT\&CK framework and constructs high-level scenario graphs (HSGs) to identify attack behaviors. RapSheet \cite{hassan2020tactical} also designs ATT\&CK behavior matching rules to construct tactical provenance graphs (TPGs) for identifying causally related threat behavior sequences. POIROT \cite{milajerdi2019poirot} constructs query graphs based on the correlation of CTI and models threat detection as a graph matching problem between query graphs and provenance graphs. APTSHIELD \cite{zhu2023aptshield} designs process and file labels and their propagation rules to achieve real-time attack response and alerting. Sleuth \cite{hossain2017sleuth} employs label-based attack detection and reconstruction techniques. Unlike these systems, OMNISEC does not require fixed rule templates and can detect unknown threats by utilizing the information in the provenance graph.

\textbf{Learning-based Instrusion Detection.} These methods can generally be divided into supervised learning and unsupervised learning. Supervised learning methods identify the characteristics of attack behaviors by training on large amounts of labeled data. ATLAS \cite{alsaheel2021atlas} trains sequence-based learning models to establish key patterns of attack and non-attack behaviors, identifies nodes with attack symptoms, and thus discovers more attack nodes. TREC \cite{lv2024trec} uses neural networks for few-shot learning to identify APT tactics/techniques.Unsupervised learning methods identify abnormal events that deviate from the distribution by modeling normal system behavior. Kairos \cite{cheng2024kairos} constructs an intrusion detection framework for full-system provenance data from four dimensions: scope, attack-agnostic, timeliness, and attack reconstruction, and is capable of detecting and reconstructing complex attacks without any prior knowledge. Flash \cite{rehman2024flash} effectively encodes the semantic and structural attributes of nodes by integrating a graph neural network-based context encoder and Word2Vec embeddings, generating high-quality node embeddings. ThreaTrace \cite{wang2022threatrace} detects host-based node-level threats through a multi-model framework based on GraphSAGE. However, as shown in Section \ref{sec: 5.2}, these methods have certain limitations in terms of detection granularity and label management, which can still lead to false positives.

\textbf{Applications of LLM in the field of security.} RAAD-LLM \cite{russell2025raad} proposes an adaptive anomaly detection framework that integrates RAG with large language models, capable of dynamically adapting to changes in normal states without fine-tuning and enhancing anomaly detection performance in time series by incorporating multimodal semantic information. ShieldGPT \cite{wang2024shieldgpt} presents an LLM-based DDoS mitigation framework that can generate interpretable attack detection and response strategies. CmdCaliper \cite{huang2024cmdcaliper} introduces a command-line semantic embedding model that detects malicious commands by combining pairs of similar commands generated by LLMs. Patil et al. \cite{patil2024leveraging} use LLMs to detect zero-day vulnerabilities in cloud networks. APT-LLM \cite{benabderrahmane2025apt} captures anomalous behavioral patterns by combining LLMs with autoencoders for semantic embedding modeling. LocalIntel \cite{mitra2025localintel} leverages large language models in conjunction with global threat intelligence and local knowledge bases to generate customized threat intelligence and mitigation strategies for organizations. Unlike the aforementioned works, \sn{} integrates large language models with external knowledge bases. Through a retrieval-enhanced prompting mechanism, it not only enables anomaly detection in the provenance graph but also supports the automated reconstruction and tracking of attack paths.

\section{Conclusion}\label{sec:conclusion}
Addressing the issues of poor generalization ability, high false alarm rates, and high costs of manual analysis in existing PIDSes under complex scenarios, this paper proposes \sn{}, an intrusion detection system that integrates LLMs and external security knowledge bases. \sn{} achieves precise identification of system abnormal behaviors and automatic reconstruction of attack paths by combining lossless compression, knowledge base construction, and retrieval-enhanced prompting mechanisms. By introducing the reasoning capabilities of LLMs and joint modeling with external knowledge, \sn{} significantly improves detection performance while ensuring interpretability. Experimental results show that \sn{} achieves performance superior to state-of-the-art methods on public benchmark datasets.

\section{Acknowledgment}\label{sec:Acknowledgment}
This work is supported in part by the following grants: National Natural Science Foundation of China under Grant No. U22B2028, 62002324, and 62276234. Zhejiang Provincial Natural Science Foundation of China under Grant No. LQ21F020016, and ZCLZ24F0202. The Fundamental Research Funds for the Provincial Universities of Zhejiang under Grant No. RF-A2023009. Wenzhou Key Scientific and Technological Projects under Grant No. ZG2024007. Wenzhou Basic Scientific Research Projects under Grant No. G2024033. Key Laboratory of Data Intelligence and Governance of Wenzhou City under Grant No. KLDI2025009.

\bibliographystyle{unsrt}

\bibliography{reference.bib}

\bio{}
\endbio
\endbio
\end{CJK}
\end{document}